# The current progress of the ALICE Ring Imaging Cherenkov Detector

A. Braem[1], G. De Cataldo[1,2], M. Davenport[1], A. Di Mauro[1], A. Franco[2], A. Gallas[2], H. Hoedlmoser[1], P. Martinengo[1], E. Nappi[2], G. Paic[3], F. Piuz[1], V. Peskov[1*]

[1]CERN, Geneva, Switzerland

[2]INFN-sez.di Bari, Italy

[3]Instituto de Ciencias Nucleares UNAM, Mexico

**Abstract**

Recently, the last two modules (out of seven) of the ALICE High Momentum Particle Identification detector (HMPID) were assembled and tested. The full detector, after a pre-commissioning phase, has been installed in the experimental area, inside the ALICE solenoid, at the end of September 2006. In this paper we review the status of the ALICE/HMPID project and we present a summary of the series production of the CsI photo-cathodes. We describe the key features of the production procedure which ensures high quality photo-cathodes as well as the results of the quality assessment performed by means of a specially developed 2D scanner system able to produce a detailed map of the CsI photo-current over the entire photo-cathode surface.

Finally we present our recent R&D efforts toward the development of a novel generation of imaging Cherenkov detectors with the aim to identify, in heavy ions collisions, hadrons up to 30 GeV/c.

**Keywords**: HMPID, GEM, CsI, photo-cathode, TIC, RICH, heavy-ion.

## 1. Introduction

The ALICE Collaboration is building a dedicated heavy-ion detector to exploit the unique physics potential of nucleus-nucleus interactions at LHC energies [1]. The aim is to study the physics of strongly interacting matter at extreme energy densities, where the formation of a new phase of matter, the quark-gluon plasma, is expected. One of the important elements of this detector is the HMPID, which extends, in the central region, the particle identification capabilities of the ALICE PID system (TOF, ITS and TPC) up to 5 GeV/c. It consists of an array of seven identical RICH modules each segmented into six large area (40x60 cm$^2$) CsI photo-cathodes (PCs), providing in total 11 m$^2$ of photo-sensitive area for the detection of Cherenkov light.

The progress in the development and construction of the first five modules were reported in recent articles [2-4]; in this paper we focus on the final stage: production and testing of the last two modules.

Finally we introduce our ideas concerning possible R&D activity oriented toward the development of a novel generation of Cherenkov detectors, able to perform, in heavy-ion collisions, hadrons identification up to 30 GeV/c.

---

[*] Corresponding author: Vladimir.Peskov@cern.ch

## 2. Particular aspects of the manufacturing of the HMPID photo-cathodes

The CsI photo-cathodes are a key element of the ALICE HMPID detector and we made special efforts in order to assess their quality, before and after mounting on the MWPC, and to monitor their stability over time. Each substrate (see [5, 6] for details) before the CsI coating went through planarity measurements. The planarity threshold for acceptance was set to 125 μm. They were then cleaned in the surface treatment facility of TS-MME. The treatment consisted of an ultrasonic bath filled with NGL 17.4 ALU de-greasing fluid (50ºC), followed by rinsing with de-ionized water followed by immersion in an ultrasonic bath filled with de-ionized water (30ºC) and finally rinsed with de-natured ethyl alcohol and dried in an oven at 60ºC. They were then mounted on the wire chamber and the nominal high voltage was applied to check for any electrical or mechanical problems. The substrates were kept under HV at least during two days. After such a test, the substrates were qualified for the coating with CsI.

Before installation in the evaporation plant the substrate was kept one night in a clean room oven at 60º. Immediately prior to installation, the substrate was cleaned with reagent grade ethyl alcohol. In order to be able to assess the photo-cathode quality we upgraded the evaporation plant by adding a UV scanner system, described in detail in [4]. The scanner was able to measure the photo-current from any point of the photo-cathode surface and compare it with a reference photo-multiplier. After cross-checks with Cherenkov light emitted by charged particles in a test beam experiment, we have been able to define a photo-current acceptance level corresponding to a detected photon yield matching the physics requirements of the detector, i.e. a Cherenkov angle resolution of 3 mrad. This cross-check was especially important because the photo-current is measured in vacuum while the photo-cathodes have to operate under gas ($CH_4$). Moreover the deuterium lamp used as UV source has a photon energy spectrum quite different from the Cherenkov light to be detected.

The same set-up was also used to measure the photo-current some time after the photo-cathodes production to check their stability and long-term behavior.

During the production of the first photo-cathodes we discovered what we name the "fatigue" effect: after the successful production of about 7-10 photo-cathodes (see [4] for more details) the subsequent ones were unsuccessful, delivering a photo-current below the acceptance level. We found empirically that in order to recover the photo-cathodes production a maintenance intervention was needed. The evaporation plant was carefully cleaned to remove the CsI deposited on its inner walls followed by the pumping down of the set- up at elevated temperatures (~70°C) until the vacuum nominal value of $10^{-7}$ Torr was achieved. After this step it was possible to resume the production of photo-cathodes with the requested QE.

2-1. The "fatigue" effect, a possible explanation

To our best knowledge the fatigue effect has not been observed by other groups before; perhaps because they dealt with the production of a small number of photo-cathodes of small size (typically <10x10 $cm^2$).

When the COMPASS experiment coated photo-cathodes of a similar size to ours and in the same coating facility, the scanner system was not yet operational. Therefore it was not possible to assess the photo-cathode quality just after the CsI deposition and effects observed later can be induced by the storage and manipulation of the photo-cathodes, for example during the assembly of the detector. It is also worth mentioning that our CsI area (11 $m^2$) is twice as large as that of COMPASS one (5.5 $m^2$).

In order to understand the fatigue effect we tried to correlate the photo-current with various parameters measured before and during the evaporation: vacuum level, chemical composition of the residual gas, temperature and evaporation rate [4] but we could not find any conclusive correlation. We think that the migration of residual water inside the evaporation plant [4] is the main factor influencing the quality of our photo-cathodes. Actually the effect of water on CsI is known and has been studied for a long time [7]. CsI is a highly hygroscopic component; it easily absorbs water which changes the structure of the surface layer (CsI molecules can even dissociate into $Cs^+$ and $I^-$ ions [4, 8]) and can make it more difficult for the photo-electrons to pass through its layer into the gas.

During the production of the photo-cathodes (42 + 6 spares) we observed that after ~5-7 subsequent evaporations of CsI photo-cathodes a rather thick and visible layer of CsI was formed on the inner walls of the evaporation plant, hereafter called the parasitic layer. After each evaporation, the set up was cooled down to room temperature and then exposed to air (in order to extract the photo-cathode), therefore the parasitic layer adsorbed water from the ambient air. During the subsequent evaporation the accumulated water accumulated can migrate towards the photo-cathode changing its properties.
In order to confirm this hypothesis, an upgrade of the evaporation plant is needed, aiming to improve the residual gas analysis sensitivity and to achieve a detailed map of the temperature inside the vessel.

In order to minimize the possible influence of the parasitic layer we decided to implement the following new procedure: for the measurement of the photo-current the photo-cathode has to be moved, inside the plant vessel and under vacuum, from the evaporation position (left on fig.1) to the measurement one (right on fig. 1). During the measurement phase the right half of the vessel was kept at $60^o$ C while the left half, with the parasitic layer, was cooled down to room temperature. Then during the time between two evaporations:
- the door of the evaporation vessel was kept closed;
- the vessel itself was kept at $30^o$ C, i.e. the heating was never switched off completely.

After the implementation of this procedure, the fatigue effect was no longer observed. Although the statistics are limited (19 photo-cathodes produced so far) and do not allow us to draw solid conclusions, we think it is worth reporting our experience while investigating more deeply this effect.

2.2. Testing of the last two modules with cosmic particles

As reported in [2] we tested the first five modules with test beam at the CERN SPS accelerator. Due the accelerators shutdown in 2006 it has not been possible to test the last two modules in the same way. To verify that all elements (MWPC, radiator vessels, photo-cathodes) operate properly after the final assembly, we performed tests with cosmic particles using a dedicated stand that allowed us to flip the module so as to have the radiator vessels facing the sky, a condition not foreseen during the design of the detector. As an example fig. 2 show a cosmic ray event. The data are currently under analysis but from a visual scan of the events we expect the last two modules will be as good as the first five, for which the full analysis is available, showing on average more than 20 photo-electrons/ring detected at $\beta = 1$.

**3. New developments: towards very high momentum particles ID**

Both, results from experiments at RHIC [9] and recent theoretical developments call for an extension of the PID capability above the present HMPID (5 GeV/c) limit.
In the following section we present our ideas concerning a possible option for a detector able to make hadron identification up to 30 GeV/c in heavy-ion collisions. In particular we are investigating the possibility to develop a counter based on the Threshold Imaging Cherenkov (TIC) concept. Figures 3a and 3b show the TIC scheme, while in [10] the reader can find a detailed description of it. It is worth stressing that, in the study of heavy-ion collisions, the identification of protons plays a role as important as the identification of pions and kaons. In a RICH detector, below the Cherenkov threshold for protons and above the one for kaons, protons can be identified only by the absence of signal, i.e. using the detector as a threshold counter. However a threshold counter is not well adapted to the harsh heavy-ion environment, where it is common to deal with several hundreds charged particles per square meter. The TIC concept links the particle track to a fiducial region in which to look for the photon signal or its absence. As opposite to a classical threshold counter, the TIC can also provide some imaging information as the signal is now a "blob" in the photon detector plane. In parallel to the hardware effort, we are investigating how to exploit the blob characteristics, like dimensions or charge density inside it, in order to increase the signal significance.
It may be interesting to note how, complementary considerations apply to the identification of electrons. In this case the $\beta$ of the electron is known in advance, being equal to 1; it may therefore be of interest to replace a RICH, with its very sophisticated and expensive optics, with the more simple and robust TIC design. The TIC concept also allows for a more compact layout compared to a RICH in a focusing configuration, a non negligible advantage.

One option we are actively pursuing is the replacement of the MWPC based photon detector, as described in [10], with a windowless GEM combined with a CsI photo-cathode. In this scheme the GEM are operated in the same gas used as the radiator medium [11]. However, the GEM is a rather fragile detector and for safe operation at the high gain required to detect single photoelectron with good efficiency, several GEM in cascade are needed (see Fig. 4a) [12]. Recently, we have developed a more robust version of the GEM, a thick GEM with resistive electrodes named RETGEM [13]. The sparks in this detector, if they appear at gains larger than $10^4$, are mild (because their current is limited by the electrode resistivity) and they damage neither the detector nor the front-end electronics. During the past months, in collaboration with the ICARUS group, we considerably improved the original design of RETGEM. The original graphite coating has been replaced by coating with CuO or CrO and each hole of the detector has now dielectric rims around it, allowing for reliable operation at higher gas gain, up to $10^5$ with single step amplification [14].

We also performed tests of this detector with a CsI coated cathode. We discovered that with one RETGEM coated with CsI it is possible to reach a gain of $10^5$ and with two RETGEM, operated in cascade, a gain larger than $10^6$, see Fig.4b. Note that the maximum achievable gain of two RETGEM operated in cascade only slightly depends on the gas mixture. It can therefore operate with unconventional (for a MWPC) gases, like pure Ar or Ne, although the very large cross section for electron backscattering will considerably reduce the photon yield in Cherenkov applications.

This effect can be, at least partially, compensated by the suppression of the window between the radiator and the photon detector itself. The effect of scintillation light has also to be carefully investigated.

In cooperation with the Reagent Research Center (Moscow) we finally made the first test of a RETGEMs coated with a SbC/CsI photo-cathode (some of our test are described in [15]). The prototype could operate at gains up to $10^5$ and the QE, for $\lambda$=405 nm, was found to be 3.2%. Such a detector may open new avenues in applications; for example after further improvements it could be used for the detection of light from aerogel radiators.

**4. Conclusions**

   All seven modules of the ALICE HMPID, including the 42 photo-cathodes, have been successfully assembled and pre-commissioned. The entire detector has been installed in the experimental area, inside the ALICE solenoid, at the end of September 2006. While completing the services installation and the in-situ commissioning of the detector, aiming at being ready for the very first collisions at LHC, we are investigating the possible options to extend our particle identification capability up to 30 GeV/C. Preliminary tests indicate a RETGEM detector with CsI photo-cathodes, in a Threshold Imaging Cherenkov configuration, as a promising candidate for this task, although more studies are needed in order to understand and optimize the photosensitive detector.

**Figure captions:**

Fig. 1. A schematic drawing of the CERN evaporation plant (see [4] for more details).
Fig. 2. Example of Cherenkov ring produced by cosmic ray.
Fig 3a and 3b: schematic view of the TIC concept.
Fig. 4a. Overall gains of triple GEM coated with CsI photocathodes measured in Ne and Ar at pressure of 1 atm.
Fig. 4b. Gains of double RETGEM coated with CsI photocathodes measured in various gases at pressure of 1 atm.

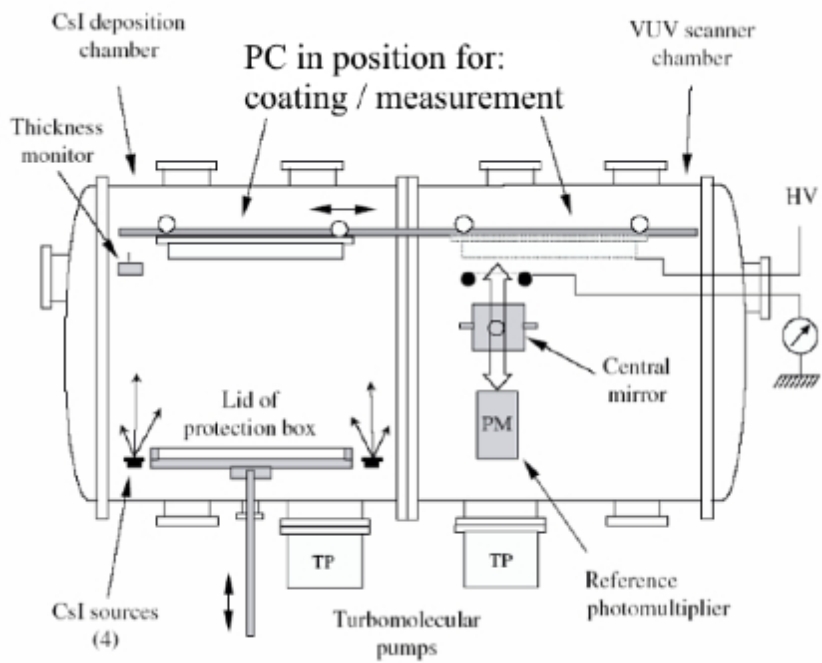

**Fig. 1**

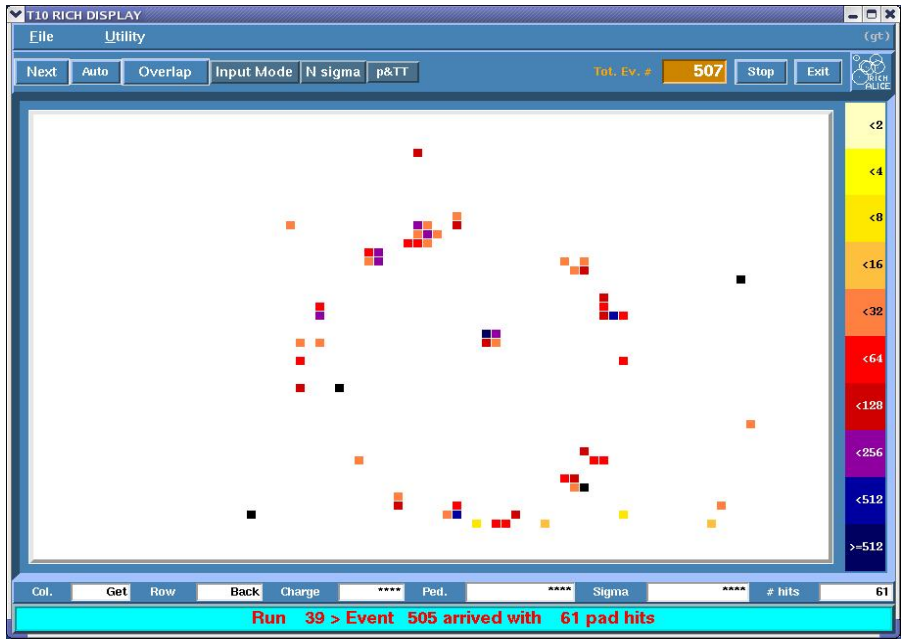

**Fig. 2**

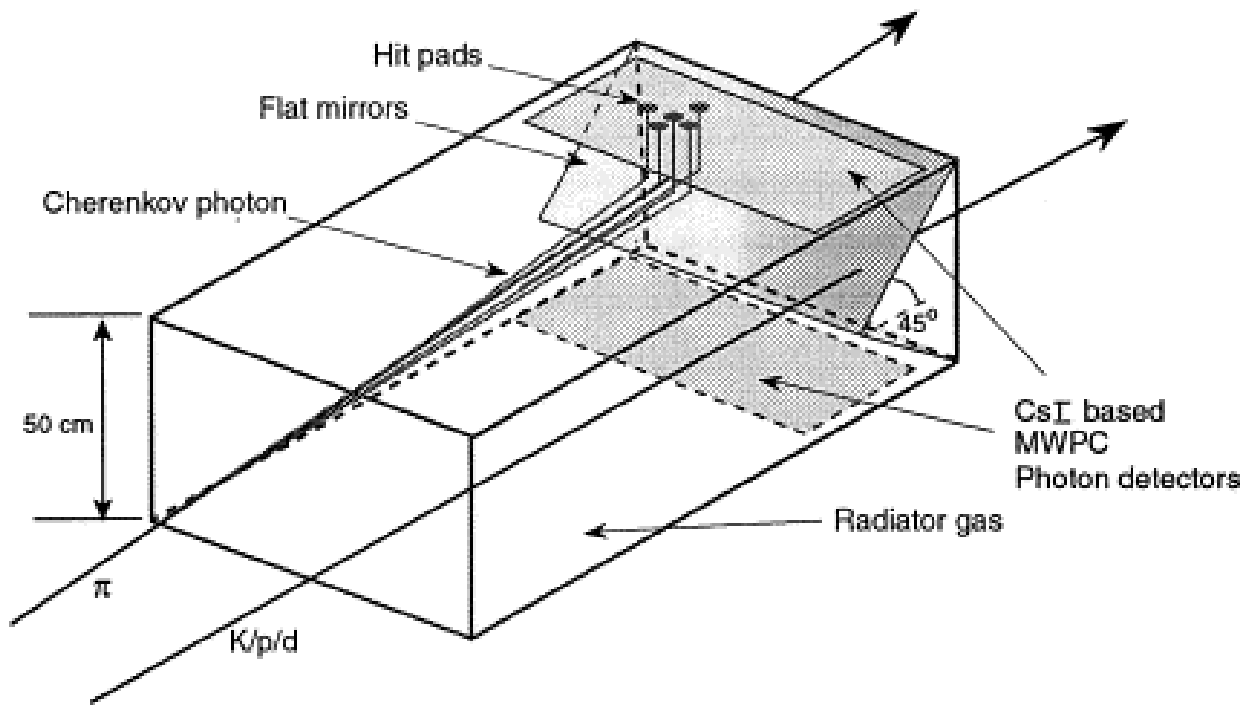

**Fig. 3a**

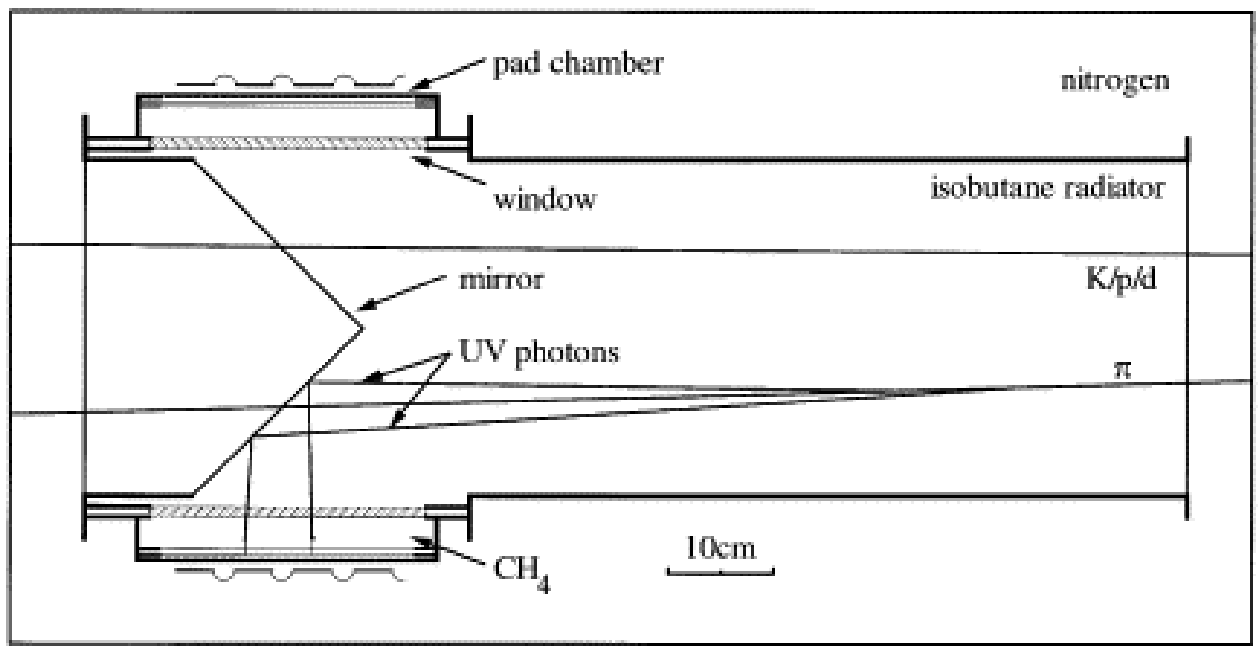

**Fig. 3b**

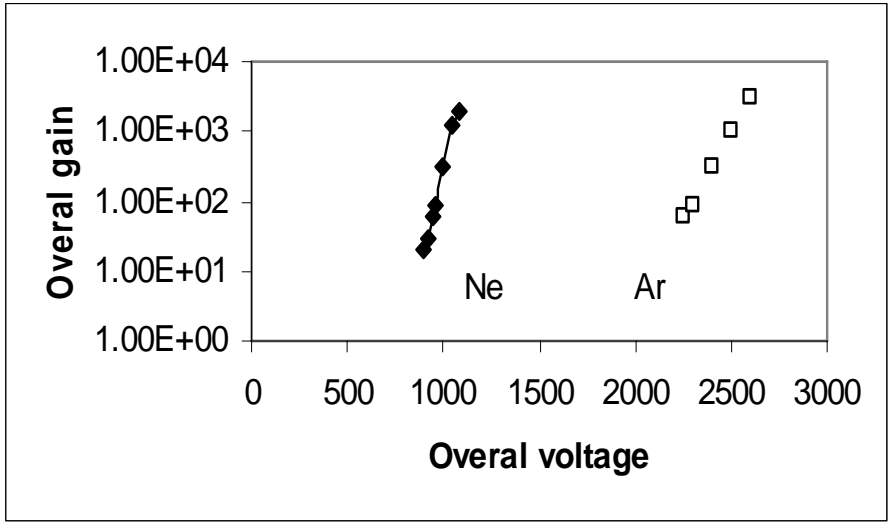

**Fig. 4a**

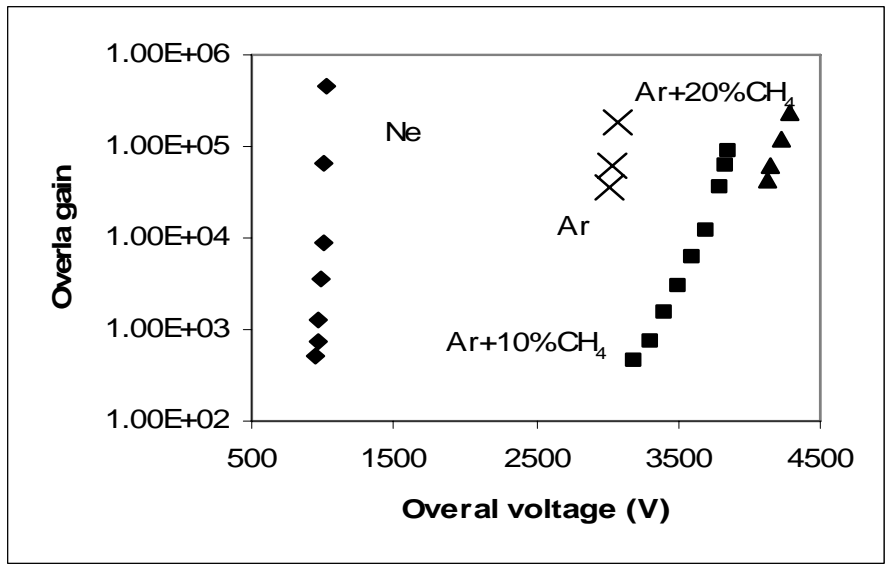

**Fig. 4b**